\documentclass[twocolumn,prb,aps,showpacs,preprintnumbers,superscriptaddress]{revtex4}
\pdfoutput=1

\usepackage{graphicx}
\usepackage{amsmath}

\newcommand{\be}{\begin{equation}}
\newcommand{\ee}{\end{equation}}
\newcommand{\bea}{\begin{eqnarray}}
\newcommand{\eea}{\end{eqnarray}}

\begin{document}

\title{Large-amplitude driving of a superconducting artificial atom: \\
Interferometry, cooling, and amplitude spectroscopy}

\author{W.~D.~Oliver}
 \email{oliver@ll.mit.edu}
 \affiliation{MIT Lincoln Laboratory, 244 Wood Street, Lexington, MA 02420}
 \affiliation{Research Laboratory of Electronics, Massachusetts Institute of Technology, Cambridge, MA 02139}
\author{S.~O.~Valenzuela}
 \email{sov@mit.edu}
 \altaffiliation[Present address: ]{ICREA and CIN2 (CSIC - ICN), Bellaterra, Barcelona, Spain.}
 \affiliation{MIT Francis Bitter Magnet Laboratory, Cambridge, MA 02139}
%


\begin{abstract}
Superconducting persistent-current qubits are quantum-coherent artificial atoms with multiple, tunable energy levels. In the presence of large-amplitude harmonic excitation, the qubit state can be driven through one or more of the constituent energy-level avoided crossings. The resulting Landau-Zener-St\"{u}ckelberg (LZS) transitions mediate a rich array of quantum-coherent phenomena. We review here three experimental works based on LZS transitions:
Mach-Zehnder-type interferometry between repeated LZS transitions, microwave-induced cooling, and amplitude spectroscopy.
These experiments exhibit a remarkable agreement with theory, and are extensible to other solid-state and atomic qubit modalities. We anticipate they will find application to qubit state-preparation and control methods for quantum information science and technology.
\end{abstract}
\pacs{03.67.Lx, 03.65.Yz, 07.60.Ly, 39.25.+k, 85.25.Cp, 85.25.Dq}
\maketitle
\vspace{-10mm}

\section{Introduction}
\label{intro}
Superconducting qubits are solid-state artificial atoms, comprising lithographically defined Josephson tunnel junctions and superconducting interconnects. When cooled to milli-Kelvin temperatures, these qubits exhibit quantized states of charge, flux, or junction phase depending on the circuit design parameters~\cite{Clarke88,Clarke08a}. Associated with these quantized states is a spectrum of energy levels, tunable via an external control parameter, e.g., an applied electric or magnetic field. Although generally only the lowest two energy eigenstates are utilized for quantum information science applications, the energy spectrum indeed extends to higher-energy levels corresponding to higher-excited states of the circuit. The separation between pairs of energy levels typically falls in the radio frequency and microwave regimes, and resonantly driving the artificial atom with a harmonic field can couple and induce quantum-state transitions.

Due to their relatively large size, superconducting artificial atoms can be strongly coupled to their external control fields. It is this feature, along with their quantum coherence, that we utilize in the present article. A large-amplitude harmonic control field can drive an artificial atom throughout its energy-level spectrum. When driven through an avoided level crossing, a Landau-Zener-St\"{u}ckelberg (LZS) transition occurs. This is a coherent process akin to a beamsplitter for photons, taking an input state of the atom and outputting a superposition of states. Repeated passages through an avoided crossing act as an atom interferometer, causing the atomic superposition state to interfere quantum mechanically with itself. Since the weighting of the superposition state depends sensitively on the size of the avoided crossing and the velocity (change in relative energy between levels per unit time) with which it is traversed, the quantum interference reflects the energy spectrum of the artificial atom. In turn, the quantum interference can be leveraged to facilitate non-adiabatic quantum control.

We begin this article with an introduction to the superconducting persistent-current qubit~\cite{Mooij99a,Orlando99a} and an overview of Landau-Zener-St\"{u}ckelberg (LZS) transitions. We then present three experimental works that utilize LZS transitions in a strongly-driven persistent-current qubit. The first is Mach-Zehnder-type interferometry between repeated LZS transitions~\cite{Oliver05a}, for which we observed quantum interference fringes in n-photon transition rates~\cite{Berns06a}, with n=1 \ldots 50. The second is microwave-induced cooling~\cite{Valenzuela06a}, by which we achieved effective qubit temperatures less than 3 mK, a factor 10-100 times lower than the environmental temperature. The third is amplitude spectroscopy~\cite{Berns08a,Rudner08a}, a spectroscopy technique that monitors the system response to amplitude rather than frequency. Amplitude spectroscopy allowed us to probe the energy spectra of our artificial atom from 0.01 - 120 GHz, while driving it at a fixed frequency near 0.2 GHz.
Finally, we conclude by considering the application of LZS transitions to quantum information science and technology.

\section{Persistent-Current Qubit: Superconducting Artificial Atom}
\label{sec:PCQB}
Superconducting artificial atoms exhibit a high degree of quantum coherence, and there have been numerous proposals and demonstrations of quantum phenomena in these systems, many derived from the fields of atomic physics and quantum optics. A few examples include: coherent superpositions of macroscopic states~\cite{Friedman00a,Wal00a,Berkley03a}, Rabi oscillations~\cite{Nakamura99a,Nakamura01,Vion02a,Yu02a,Martinis02a,Chiorescu03a,Claudon04a,Plourde05a,Saito06a,Lisenfeld07a}, Landau-Zener transitions~\cite{Izmalkov04a}, St\"{u}ckelberg oscillations~\cite{Oliver05a,Berns06a,Sillanpaa06a,Wilson07a,Izmalkov08a},
microwave cooling~\cite{Valenzuela06a,Niskanen07b,You08a,Grajcar08a}, electromagnetically induced transparency~\cite{Murali04a,Dutton06a}, geometrical phase~\cite{Leek07a}, and cavity quantum electrodynamics~\cite{Chiorescu04a,Wallraff04a,Johansson06a,Deppe08a,Fink08a,Fragner08a,Hofheinz08a}. Significant progress has also been made toward their application to quantum information science~\cite{Makhlin01a,Mooij05}, including state initialization~\cite{Valenzuela06a}, tunable~\cite{Hime06a,Ploeg07a,Niskanen07a,Kerman08a} and long-distance~\cite{Sillanpaa07a,Majer07a} coupling, quantum control~\cite{Pashkin03a,Yamamoto03a,McDermott05a,Plantenberg07a}, quantum state~\cite{Steffen06a,Steffen06b} and process~\cite{Neeley08a} tomography, and measurement~\cite{Siddiqi04a,Katz06a,Lupascu07a,Yamamoto08a,Beltran08a,Naaman08a}.
For a recent review of superconducting qubits, see Ref.~\onlinecite{Clarke08a}.

\begin{figure*}
    \begin{center}
  \includegraphics[width=5in]{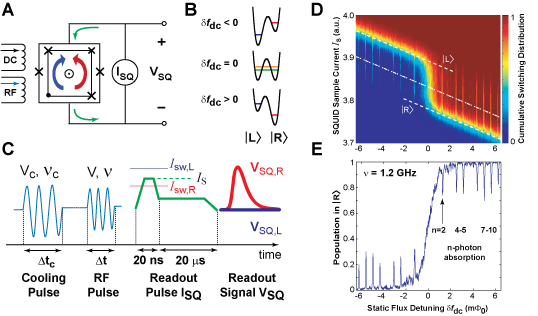}
\caption{Artificial atom (persistent current qubit) and measurement set-up. A Schematic of the qubit and surrounding DC SQUID readout. B Double well qubit potential comprising energy levels for static magnetic flux bias $\delta f_{\textrm{dc}}$ about $\Phi_0/2$, where $\Phi_0$ is the superconducting flux quantum. Diabatic states of the left (right) well corresponds to a persistent current with clockwise (counterclockwise) circulation. At detuning $\delta f_{\textrm{dc}}=0$, the double-well potential is symmetric and the diabatic-state energies are degenerate. Tunnel coupling opens an avoided crossing $\Delta$. C Qubit excitation and read-out pulse sequence. The qubit is first prepared in its ground state with a harmonic cooling pulse with amplitude $V_{\textrm{c}}$ and frequency $\nu_{\textrm{c}}$. Quantum-state transitions are induced with a subsequent harmonic RF pulse with amplitude $V$ and frequency $\nu$. The qubit state is read-out using the DC SQUID switching response. D Qubit step. Cumulative switching current distribution of the SQUID for each $\delta f_{\textrm{dc}}$ value following a 3-$\mu$s RF driving pulse at 1.2 GHz applied to the qubit (the cooling pulse was not used here). Resonant multiphoton transitions (of order n) are observed between states $|L \rangle$ and $|R \rangle$. The switching distribution along the dashed-dotted line discriminates between states $|L \rangle$  and $|R \rangle$ (E).}
\label{fig:1}       
    \end{center}
\end{figure*}

Here, we use a superconducting persistent-current qubit to implement an artificial atom~\cite{Mooij99a,Orlando99a}. The persistent-current qubit is a superconducting loop interrupted by three Josephson junctions (Fig.~\ref{fig:1}A). When biased with a static magnetic flux $f_{\textrm{dc}} \sim \Phi_0/2$, where $\Phi_0$ is the superconducting flux quantum, the system assumes a double-well potential profile (Fig.~\ref{fig:1}B). The diabatic ground state of the left (right) well corresponds to a persistent current $I_{\textrm{q}}$ with clockwise (counterclockwise) circulation. These two diabatic energy levels (Fig.~\ref{fig:2}A) have a separation $\varepsilon = 2I_{\textrm{q}} \delta f_{\textrm{dc}}$ linear in the flux detuning $\delta f_{\textrm{dc}} \equiv f_{\textrm{dc}} - \Phi_0/2$. Higher-excited states of the double-well potential (see Fig.~\ref{fig:5}C) will be considered in Sections~\ref{sec:Cooling} and~\ref{sec:AS}.

The two-level Hamiltonian for the lowest two diabatic states is shown in Eq.~\ref{Eq:Hamiltonian}. At detuning $\delta f_{\textrm{dc}}=0$, the double-well potential is symmetric and the diabatic-state energies are degenerate. At this ``degeneracy point,'' resonant tunneling between the diabatic states opens an avoided level crossing of energy $\Delta$. Here, the qubit states are $\sigma_x$ eigenstates, corresponding to symmetric and anti-symmetric combinations of diabatic circulating-current states. Detuning the flux away from this point tilts the double well, allowing us to tune the eigenstates and eigenenergies of the artificial atom. Far from the degeneracy point the qubit states are approximately $\sigma_z$ eigenstates, the diabatic states with well-defined circulating current. The qubit is read out using a hysteretic DC SQUID (superconducting quantum interference device), a sensitive magnetometer that can distinguish the flux generated by circulating current states.

In addition to the static flux biases, the artificial atom is controlled and read out using the pulses illustrated in Fig.~\ref{fig:1}C. As we describe below, the qubit is first prepared in its ground state using a harmonic cooling pulse with amplitude $V_{\textrm{c}}$ and frequency $\nu_{\textrm{c}}$. Quantum-state transitions are then driven using a harmonic RF pulse with amplitude $V$ and frequency $\nu$. These fields are mutually coupled to the qubit through a small antenna. This is followed by a SQUID readout current pulse using the ``sample and hold'' technique~\cite{Oliver05a,Chiorescu03a}. If the sample current exceeds the SQUID switching current, a voltage pulse will appear at the output during the hold phase. Threshold detection looks for the presence or absence of a SQUID voltage, and this constitutes a digital measurement of the qubit state. Alternatively, although not used in these experiments, we have incorporated the SQUID into a resonant circuit and realized qubit readout via the shift in resonance frequency and phase for both the linear and non-linear resonance regimes~\cite{Lee05a,Lee07a}.

The ``qubit step'' is shown in Fig.~\ref{fig:1}D as a function of the SQUID sample current and the flux detuning. The diabatic states $|L \rangle$ and $|R \rangle$ correspond to different levels of sample current (dashed lines) located symmetrically about the degeneracy point $\delta f_{\textrm{dc}}$=0. This plot constitutes a cumulative switching current distribution of the SQUID for each $\delta f_{\textrm{dc}}$ value. Additionally, a 3-$\mu$s RF pulse at 1.2 GHz is applied to the qubit, and resonant transitions can be observed as fingers extending down (up) from state  $|L \rangle$ ($|R \rangle$) when $\textrm{n}\times$1.2 GHz becomes resonant with the energy-level separation. A best-estimator (dashed-dotted line) can be determined to provide the best statistical  discrimination between states  $|L \rangle$  and $|R \rangle$. The resulting qubit step with its saturated n-photon resonances along the best estimator line is shown in Fig.~\ref{fig:1}E.

The device used in this work was fabricated at Lincoln Laboratory using a fully-planarized niobium trilayer process and optical lithography. It has a critical current density $J_{\rm c} \approx 160 \,{\rm A/cm^2}$, and the characteristic Josephson and charging energies are $E_{\textrm{J}} \approx (2\pi\hbar)300\,{\rm{GHz}}$ \,\,{\textrm{and}}\,\, $E_{\textrm C} \approx (2\pi\hbar)0.65\,{\textrm{GHz}}$ respectively. The ratio of the qubit Josephson junction (JJ) areas is $\alpha \approx 0.84$, and $\Delta \equiv \Delta_{0,0} \approx (2\pi\hbar)10\,{\textrm{MHz}}$. Although dependent on the flux detuning, the approximate values for coherence times are: interwell relaxation time $T_1 \sim 100 \,\mu \textrm{s}$, intrawell relaxation time $T_1 \sim 0.05 \,\mu \textrm{s}$, homogeneous transverse decay time $T_2 \sim 20 \, \textrm{ns}$, inhomogeneous transverse decay time $T_2^* \sim 10 \, \textrm{ns}$. The experiments were performed in a dilution refrigerator at a base temperature of 20 mK. The device was magnetically shielded, and all electrical lines were carefully filtered and attenuated to reduce noise (see Ref.~\onlinecite{Berns08a} for details).

\section{Landau-Zener-St\"{u}ckelberg Transitions}
\label{sec:LZS}
Landau-Zener-St\"{u}ckelberg (LZS) transitions occur when a quantum system is driven through an energy-level avoided crossing~\cite{Landau32a,Zener32a,Stueckelberg32a}. The resulting quantum dynamics of the LZS mechanism~\cite{{H_Nakamura01a}} in driven systems~\cite{Grifoni98a} have been developed~\cite{Kayanuma93a,Kayanuma94a} within a two-level coherent scattering formalism~\cite{Kayanuma97a,Kayanuma00a,Shytov03,Ashab07a} with potential application to quantum information science and technology~\cite{Shytov03,Ashab07a,Wubs06a,KSaito06a,KSaito07a}.

\begin{figure*}
\begin{center}
  \includegraphics[width=5in]{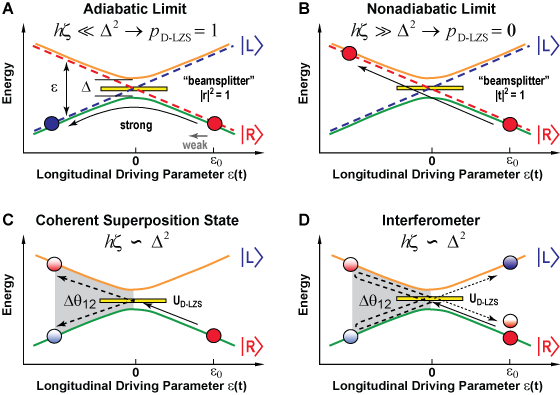}
\caption{Landau-Zener-St\"{u}ckelberg transition. A Adiabatic limit: $h \zeta \ll \Delta^2$; the probability of a transition from state $| R \rangle$ to state $| L \rangle$ approaches unity. B Nonadiabatic limit: $h \zeta \gg \Delta^2$; the probability of a transition from state $| R \rangle$ to state $| L \rangle$ approaches zero. C Coherent superposition of states. At intermediate sweep velocities ($h \zeta \sim \Delta^2$) a superposition state $\alpha | L \rangle + \beta | R \rangle$ results from an excursion through the avoided crossing. A phase difference $\Delta \theta_{12}$ accrues due to the energy difference between the states $| L \rangle$ and $| R \rangle$. D Quantum interference during a second excursion through the avoided crossing. Interference fringes appear at half-integer and integer values of $\Delta \theta_{12}/2 \pi$, which is tunable via the drive amplitude.}
\label{fig:2}       
\end{center}
\end{figure*}

We model a LZS transition at a single crossing using the two-level Hamiltonian
\begin{equation}
 \mathcal{H} =
    -\frac{1}{2} \left(
    \begin{matrix}
        \varepsilon(t) & \Delta \\
        \Delta & -\varepsilon(t)
    \end{matrix}
    \right),
 \label{Eq:Hamiltonian}
\end{equation}
in which $\varepsilon$ is the energy difference between diabatic-state energy levels (dashed lines in Fig.~\ref{fig:2}a), and $\Delta$ is the size of the avoided crossing and corresponds to the coupling strength between diabatic states $| \textrm{L} \rangle$ and $| \textrm{R} \rangle$.


The system is prepared in state $| R \rangle$ at a static value $\varepsilon_0 \gg \Delta$ far from the avoided crossing (blue dot, Fig.~\ref{fig:2}A) and driven longitudinally from that point by making $\varepsilon(t)$ time dependent. We distinguish here between ``weak driving,'' in which the resulting excursion is not large enough to reach the avoided crossing, and ``strong driving'' for which the avoided crossing is traversed. Under strong-driving conditions, the asymptotic probability $P_{\textrm{D-LZS}}$ of a transition between diabatic states,
\begin{equation}
 P_{\textrm{D-LZS}} \equiv 1 - P_{\textrm{E-LZS}} = 1 - \exp{
        \frac{- \pi \Delta^2}{2 \hbar \zeta_i}},
 \label{Eq:P_LZS}
\end{equation}
is governed by the relative-energy sweep rate $\zeta_i$
\begin{equation}
 \zeta_i = \frac{d\varepsilon(t)}{dt} |_{t=t_i}
\end{equation}
evaluated at the time $t=t_i$ at which the system is swept through the avoided crossing. In the original LZS formulation, the system was driven with a fixed sweep rate, 
whereas we will later consider a harmonically driven system.
We also note that we have elected to monitor the probability $P_{\textrm{D-LZS}}$ of a transition between the diabatic states $| L \rangle$ and $| R \rangle$, because our readout detector is sensitive to changes in diabatic state. This can be written with no loss of generality in terms of the probability $P_{\textrm{E-LZS}}$ of a transition between the system eigenstates.

There are two strong-driving limits characterized by the relative sizes of the sweep velocity $\zeta$ and the avoided crossing $\Delta$. In the adiabatic limit (Fig.~\ref{fig:2}A), the sweep velocity is small ($h \zeta \ll \Delta^2$) and the probability of a transition from state $| R \rangle$ to state $| L \rangle$ approaches unity, $P_{\textrm{D-LZS}}\rightarrow 1$. In this case, the system dynamics are slow enough that the system adiabatically follows the ground eigenstate through the avoided crossing. In the nonadiabatic limit (Fig.~\ref{fig:2}B), the sweep velocity is large ($h \zeta \gg \Delta^2$) and the probability of a transition approaches zero $P_{\textrm{D-LZS}}\rightarrow 0$. In this case, the dynamics are too fast for the system to follow; the system remains in diabatic state $| R \rangle$ and, thereby, effectively jumps the energy gap at the avoided crossing.

More generally, a superposition state $\alpha | L \rangle + \beta | R \rangle$ results from an excursion through the avoided crossing, as illustrated in Fig.~\ref{fig:2}C.
Following an idea discussed by Shytov {\it et al.}~\cite{Shytov03}, the LZS transition acts as a beamsplitter for the atomic state.
The amplitudes $\alpha$ and $\beta$ are determined by a unitary transformation $U_{\textrm{D-LZS}}$, effectively a $2 \times 2$ ``beamsplitter matrix'' comprising complex reflection $r$ and transmission $t$ coefficients related to the adiabaticity parameter~\cite{H_Nakamura01a} $\Delta^2 / \hbar \zeta$
present in Eq.~\ref{Eq:P_LZS} such that $|t|^2= 1-P_{\textrm{D-LZS}}$, $|r|^2= P_{\textrm{D-LZS}}$, and  $U^{\dagger}_{\textrm{D-LZS}} U_{\textrm{D-LZS}} = I$. Note that we have defined $|t|^2$ and $|r|^2$ from the perspective of a beamsplitter, which ``transmits'' (``reflects'') an input state to the same (opposite) diabatic state, respectively.
After the transition, a relative phase $\Delta \theta_{12}$ accrues due to the energy difference between the states $| L \rangle$ and $| R \rangle$. If the drive $\varepsilon(t)$ then returns the system through the avoided crossing a second time, the atomic state collides and quantum mechanically interferes with itself during the second LZS transition (Fig.~\ref{fig:2}D). The cumulative result is an atom-state interferometer whose output state depends on the LZS transition amplitudes and the interference phase.

An analogy can be made to an optical Mach-Zehnder interferometer: the atomic states play the role of the photon modes, the LZS transitions play the role of the photon beamsplitters, and the energy-level splitting, which determines the interference phase, plays the role of optical path length difference. The quantum interference is robust provided the evolution time of the state through the interferometer is short compared with the atom's coherence times. In addition to superconducting artificial atoms~\cite{Oliver05a,Berns06a,Nakamura01,Sillanpaa06a,Wilson07a,Saito06a}, this concept is generally applicable to other solid-state artificial atoms~\cite{vanderWiel03a,Hanson07a} and generalized spin systems~\cite{Tannoudji92a} (e.g., molecular magnets~\cite{Friedman96a,Thomas96a,Wernsdorfer99a}, natural atoms~\cite{Baruch92a,Yoakum92a}, and molecules~\cite{Mark07a,Mark07b,Lang08a}) that exhibit avoided level crossings, and it is extensible to multiple energy levels as we demonstrate in Section~\ref{fig:5}.

\section{Mach-Zehnder-type Interferometry}
\label{sec:MZ}
The structure of the n-photon spectroscopy peaks seen in Fig.~\ref{fig:1}E consists of regularly-spaced resonances positioned according to the condition $\textrm{n}\times$1.2 GHz being resonant with the energy level separation. Notably, however, for this particular value of driving amplitude, the n=1, 3, and 6 photon peaks are missing. As we describe in this section, the presence and absence of these peaks arise from Mach-Zehnder-type quantum interference at a level crossing. The interference phase is tunable via the driving amplitude, leading to a ``Bessel staircase'' interference pattern in the observed spectroscopy. The interference oscillations are known as St\"{u}ckelberg oscillations~\cite{Stueckelberg32a}, and they have been observed in both artificial~\cite{Oliver05a,Berns06a,Sillanpaa06a,Wilson07a} and natural~\cite{Baruch92a,Yoakum92a,Mark07a} atomic systems.

\begin{figure*}
\begin{center}
  \includegraphics[width=5in]{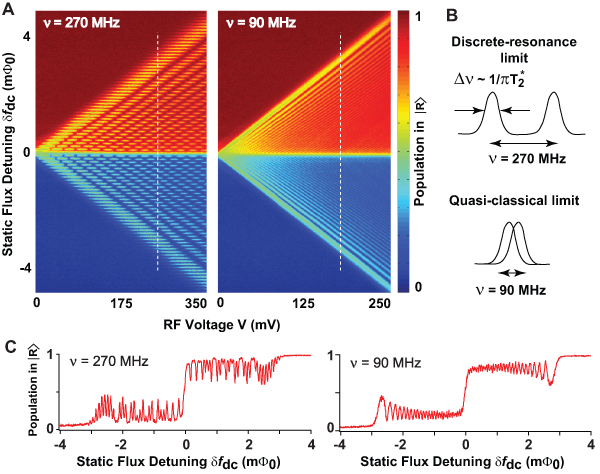}
\caption{Mach Zehnder interference. A Measured qubit population as a function of excitation amplitude and flux detuning in two excitation regimes. Left: $\nu=270$ MHz. The resonance linewidth is smaller than the excitation frequency (B, top); individual n-photon resonances can be resolved and a Bessel staircase is observed. Right: $\nu=90$ MHz. The resonance linewidth is larger than the excitation frequency (B, bottom); individual resonances are no longer resolved but coherent interference is still observed. C Interference fringes in qubit population for $\nu=270$ MHz (left) $\nu=90$ MHz (right) along the vertical dashed lines in A.}
\label{fig:3}       
\end{center}
\end{figure*}

In a conventional Mach-Zehnder interferometer, an optical signal is passed through two beamsplitters. The first beamsplitter coherently divides the signal into two output paths, which may have different effective optical path lengths. These paths are then recombined at the second beamsplitter, where the signal waves interfere and exit the interferometer through the two output ports. An intensity measurement at either output port exhibits interference fringes depending on the relative path length.

Here, we instead utilize LZS transitions at level crossings as beamsplitters for the atomic state~\cite{Shytov03}. We drive the persistent-current qubit with a harmonic driving field of the form
\begin{equation}
 \varepsilon(t) = \varepsilon_0 + A \sin \omega t
\end{equation}
with $\omega = 2\pi \nu $ the driving frequency and $A>\varepsilon_0$ the field amplitude (in units of energy), which is proportional to the RF voltage at the source.
As illustrated in Fig.~\ref{fig:2}D, the qubit state undergoes two LZS transitions during one period of the driving field. The first LZS transition at time $t_1$ splits the qubit into a superposition of excited and ground states. A relative phase $\Delta \theta_{12}$ accumulates
\begin{equation}
  \Delta\theta_{12}=\frac1{\hbar}\int_{t_1}^{t_2} \epsilon(t)dt ,\quad \epsilon(t)=\epsilon_{|R\rangle}(t)-\epsilon_{|L\rangle}(t)
\end{equation}
until the second LZS transition at time $t_2$, at which point the qubit state collides with itself and interferes. Interference fringes appear at half-integer and integer values of $\Delta\theta_{12}/2 \pi$, which is tunable via the drive amplitude.

It is clear, however, that a second phase must also play a role in this problem, since the qubit state continues to evolve for the remainder of the driving period. It is physically meaningful to consider the total phase $\theta$ accumulated over a single period:
\begin{equation}
 \theta = \frac1{\hbar} \oint \epsilon(t)dt=2\pi\epsilon_0/\hbar\omega,
\end{equation}
which is {\it independent} of the driving amplitude. Over many periods of the driving field, the cascaded pairs of LZS transitions (cascaded interferometers) will constructively interfere provided $\theta = 2\pi n$. One can view this as a ``time-domain'' formulation for the familiar n-photon resonance condition
\begin{equation}
 \varepsilon_{0,n} = n h \nu
\end{equation}
where n is the number of photons involved in the transition. It is only when the product $n h \nu$ equals the energy separation $\varepsilon_{0,n}$ that the cascaded LZS transitions lead to a net buildup of state population and, as a function of $\Delta \theta_{12}$, the observed interference fringes. These oscillations are related to photoassisted transport~\cite{Tien63a,Kouwenhoven94,NakamuraTsai99} and Rabi oscillations~\cite{Nakamura01,Saito06a} in the multiphoton regime.

Mach-Zehnder-type interference in the discrete-resonance limit driven towards saturation is shown in Fig.~\ref{fig:3}A for driving frequency $\nu=270$ MHz. This frequency is larger than the resonance linewidth (Fig.~\ref{fig:3}B), and so individual n-photon resonances can be resolved. There are two main features observable in this plot. The first is the presence of equally-spaced n-photon transitions as a function of flux detuning, symmetrically located about the qubit step at $\delta f_{\textrm{dc}}=0$. As one might expect, the onset of the higher-photon transitions requires larger driving amplitudes. Remarkably, we observe up to 50-photon transitions in this scan. The second main feature is that for each n-photon resonance, the spectroscopy appears and disappears as a function of amplitude, which sweeps the interference phase $\Delta \theta_{12}$. For example, 14 oscillation lobes are visible for the 1-photon transition. A vertical slice of the spectroscopy (dashed white line in Fig.~\ref{fig:3}A) is plotted in Fig.~\ref{fig:3}C; at this particular amplitude, one can see the enhancement and suppression of the spectroscopy peaks as was observed in Fig.~\ref{fig:1}E.

The Mach-Zehnder interference for an n-photon transition yields a modified amplitude-dependent matrix element~\cite{Oliver05a,Nakamura01,Tannoudji92a}
\begin{equation}
 \Delta_n = \Delta J_n(\lambda)
\end{equation}
where $J_n(\lambda)$ is the nth-order Bessel-function, and its argument $\lambda = A / h \nu$ is the dimensionless driving amplitude. Intuitively, the Bessel-function dependence arises, because the qubit is driven harmonically through energy levels that are linear in flux detuning and, as a result, the interference phase accumulates with a harmonic time dependence. The transition rate and, therefore, the population transfer, at each n-photon resonance is related to the matrix element squared, $\Delta_n^2$. At specific amplitudes, despite driving the system resonantly, no net transition occurs due to a complete destructive Mach-Zehnder interference corresponding to the zeros of $J_n(\lambda)$; this is the coherent destruction of tunneling~\cite{Grossmann91a,Llorente92a,Kayanuma08a} condition for driven n-photon transitions~\cite{Oliver05a,Berns06a}. More generally, as a function of the driving amplitude, we have observed the continuous evolution between regions of enhanced and suppressed tunneling as dictated by $J_n(\lambda)$ over several Bessel-function lobes (Fig.~\ref{fig:3}A).

In the discrete resonance limit, n-photon resonances are distinguishable, because the coherence time of the qubit is sufficiently longer than the driving period. By reducing the frequency to $\nu = 90$ MHz (Fig.~\ref{fig:3}A, right panel), we effectively made the drive period comparable with the linewidth, $\nu T_2^* \sim 1$ (Fig.~\ref{fig:3}B). In this spectroscopy plot, the individual n-photon resonances are no longer resolvable. However, the Mach-Zehnder interference fringes (vertical slice, Fig.~\ref{fig:3}C) are clearly visible, because the qubit remains coherent during the critical fraction of the drive period during which the phase $\Delta \theta_{12}$ accumulates and the Mach-Zehnder interference occurs. Intuitively, provided $\nu T_1 \gg 1$, the output populations of each quantum interferometer are then preserved (frozen) until the subsequent interferometer is reached in the following period; therefore, although the resonance condition is lost, the Mach-Zehnder quantum interference remains. This behavior can be contrasted with driven Rabi oscillations, for which there would be no signature of coherence in the limit $\nu T_2^* \sim 1$.
%

\begin{figure*}
\begin{center}
  \includegraphics[width=5in]{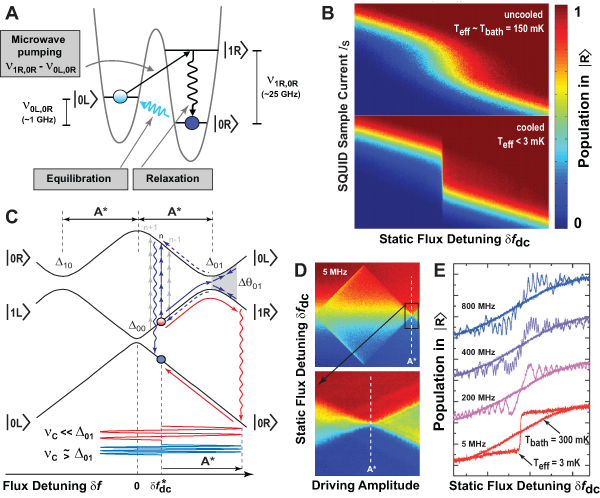}
\caption{Cooling of an artificial atom via an ancillary excited state. A External excitation transfers the thermal population from state $|0L \rangle$ to state $|1R \rangle$ (straight line) from which it decays into the ground state $|0R \rangle$. Wavy lines represent spontaneous relaxation and absorption leading to equilibration. B Qubit step at $T_{\rm bath}$ = 150 mK in equilibrium with the bath (top) and after a 3-$\mu$s cooling pulse at 5 MHz (bottom). The average level populations exhibit a qubit step about $\delta f_{\rm {dc}} = 0$, with a width proportional to $T_{\rm bath}$ (top) and $T_{\rm eff} \ll T_{\rm bath}$ (bottom). C Schematic level diagram illustrating resonant and adiabatic cooling. $|0L \rangle \to |1R \rangle$ transitions are resonant at high driving frequency $\nu$ (blue lines) and occur via adiabatic passage at low $\nu$ (red lines). $\Delta_{00}$ and $\Delta_{01}$ are the tunnel splittings between $|0R \rangle$ - $|0L \rangle$ and $|0L \rangle$ - $|1R \rangle$. D Optimal cooling parameters. State $|0R \rangle$ population vs. flux detuning $\delta f_{\rm dc}$ and driving amplitude $A$ with $\nu$ = 5 MHz, $\Delta t_c = 3\,\mu{\rm s}$, and $T_{\rm bath}$= 150 mK. Optimal conditions for cooling are realized at $A = A^*$, where $A^*$ is defined in C. E Cooling at driving frequencies $\nu$ = 800, 400, 200 and 5 MHz. State $|0R \rangle$ population vs. $\delta f_{\rm dc}$ for the cooled qubit and for the qubit in thermal equilibrium with the bath (black lines, $T_{\rm bath} = 300\,{\rm mK}$). Measurements for $\nu=$ 800, 400, 200 and 5 MHz are displaced vertically for clarity. A cooling factor of 100, independent of detuning, is obtained in the adiabatic limit (5 MHz).}
\label{fig:4}       
\end{center}
\end{figure*}

%
\section{Microwave Cooling}
\label{sec:Cooling}

%

\begin{figure*}
\begin{center}
  \includegraphics[width=5in]{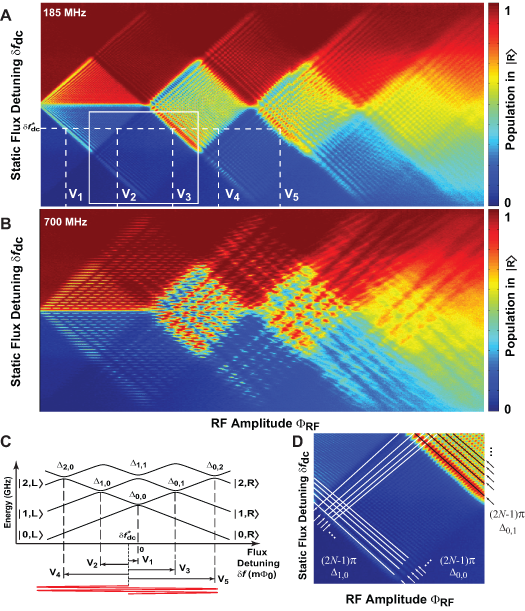}
\caption{Amplitude spectroscopy with long-pulse driving towards saturation. Spectroscopy diamonds obtained at driving frequencies $\nu=185$ MHz (A) and $\nu=700$ MHz (B). The color scale represents the net qubit population in state $|q,R\rangle$, where $R$ ($L$) labels diabatic states of the right (left) well of the qubit double-well potential, and $q=0,1,2,...$ labels the longitudinal modes. The excitation amplitude $V$ is swept for each static flux detuning $\delta f_{\rm{dc}}$. The diamond edges mark the driving amplitude $V$ for each value of $\delta f_{\rm{dc}}$ when an avoided level crossing is first reached (amplitudes $V_1-V_5$ for $\delta f_{\rm{dc}} = \delta f_{\rm{dc}}^*$). C Schematic energy-level diagram illustrating the relation between $V$ and the avoided crossing positions for $\delta f_{\rm{dc}}=\delta f_{\rm{dc}}^*$. The arrows represent the amplitudes $V_1-V_5$ in A. (D) Zoom in of 185-MHz interference patterns (box region, Fig.~\ref{fig:5}A). The arrows point to lines of constant phase $(2N-1)\pi$ along which LZS transitions are likely to occur for the avoided crossings $\Delta_{0,0}$, $\Delta_{0,1}$, and $\Delta_{1,0}$.}
\label{fig:5}       
\end{center}
\end{figure*}

The previous discussion involved driving transitions in the lowest two energy levels in the double-well potential of our artificial atom (Fig.~\ref{fig:4}A), which constitute the two-level qubit subsystem of a more complex energy level diagram (Fig.~\ref{fig:5}C).  When higher-excited states are accessed, the driven system behavior can be markedly different from the population saturation observed when only two levels are involved. For example, at least three levels are required to achieve population inversion, and such a multi-level artificial atom coupled to a microwave cavity has been used to demonstrate masing (microwave lasing)~\cite{Astafiev07a}. In that work, Josephson quasi-particle states were driven to achieve inversion. Alternatively, population inversion can be established by accessing an ancilliary excited state via direct or LZS transitions. This will be briefly discussed in the next section.

Here, 
by reversing the cycle that leads to population inversion, we show that one can pump population from the qubit excited state $|0L \rangle$ to the qubit ground state $|0R\rangle$ (Fig.~\ref{fig:4}A) via an ancillary energy level $|1R\rangle$. In the case where the population in $|0L \rangle$ results from thermal excitation, the transfer of population to $|0R\rangle$ effectively cools the qubit by lowering its effective temperature. This kind of active cooling represents a means to initialize and reset qubits with high fidelity, key elements for quantum information science and technology. Alternatively, the pumping mechanism can be used to refrigerate environmental degrees of freedom~\cite{Niskanen07b}, or to cool neighboring quantum systems~\cite{You08a,Grajcar08a}.

More explicitly, for a qubit in equilibrium with its environment, the population in $|0L\rangle$ that is thermally excited from $|0R\rangle$ follows the Boltzmann relation
\begin{equation}
 p_{0L}/p_{0R} = \exp[- \varepsilon /k_{\rm B} T_{\rm bath}],
 \label{Boltzmann}
\end{equation}
where $p_{0L,0R}$ are the qubit populations for energy levels $\varepsilon_{0L,0R}$, $\varepsilon = \varepsilon_{0L}- \varepsilon_{0R}$, $k_{\rm B}$ is the Boltzmann constant, and $T_{\rm bath}$ is the bath temperature. To cool the qubit subsystem below $T_{\rm bath}$, a microwave magnetic flux of amplitude $A$ and frequency $\nu$ targets the $|0L\rangle \to |1R\rangle$ transition, driving the state-$|0L\rangle$ thermal population to state $|1R\rangle$, from which it quickly relaxes to the ground state $|0R\rangle$. Efficient cooling only occurs when the driving-induced population transfer to $|0R\rangle$ is faster than the thermal repopulation of $|0L\rangle$. The hierarchy of relaxation and absorption rates required, $ \Gamma_{0R,1R}\gg \Gamma_{0L,1R},\Gamma_{0L,0R}$, is achieved in our system owing to a relatively weak tunneling between wells, which inhibits the interwell relaxation and absorption processes $|1R\rangle \to |0L\rangle$ and $|0R\rangle \to |0L\rangle$, compared with the relatively strong intrawell relaxation process $|1R\rangle \to |0R\rangle$.

Figure \ref{fig:4}B shows the qubit step at $T_{\rm bath}$ = 150 mK in equilibrium with the bath (top) and after a 3-$\mu$s cooling pulse at 5 MHz (bottom). Under equilibrium conditions, the average level populations exhibit a thermally-broadened qubit step about $\delta f_{\rm {dc}} = 0$, with a width proportional to $T_{\rm bath}$.
%
%
The presence of microwave excitation targeting the $|0L\rangle \to |1R\rangle$ transition, followed by relaxation, acts to increase the ground-state population and, thereby, sharpens the qubit step. Cooling can thus be quantified in terms of an effective temperature $T_{\rm eff} < T_{\rm bath}$, a signature that is evident from the narrowing of the qubit steps in Fig. \ref{fig:4}B after cooling. More precisely, using the notation from Fig.~\ref{fig:2}, the effective qubit temperature is obtained by fitting an effective temperature that would have been required in equilibrium to achieve the observed qubit population $p_{0R}$,
\begin{equation}
 p_{0R} = \frac{ \varepsilon}{\sqrt{ \varepsilon^2 + \Delta^2}}
  \left[ \tanh \left( \frac{\sqrt{ \varepsilon^2 + \Delta^2}}{2 k_{\textrm{B}} T_{\textrm{eff}}} \right)
  + 1 \right].
\end{equation}

Universal cooling (cooling that is independent of flux detuning) occurs near an optimal driving amplitude $A^*$ (Fig.~\ref{fig:4}C). This is demonstrated in Fig.~\ref{fig:4}D where we present the $|0R\rangle$ state population $P_{\rm sw}$ measured as a function of the microwave amplitude $A$ and flux detuning $\delta f_{\rm dc}$ for $\nu=5\,{\rm MHz}$. Cooling and the diamond feature can be understood in terms of the energy level diagram (Fig.~\ref{fig:5}C).
As the amplitude of the microwave pulse is increased from $V=0$, population transfer first occurs when the $\Delta_{0,0}$ avoided crossing is reached, {\it i.e.}  $A>|\delta f_{\rm {dc}}|$; this defines the front side of the observed diamond, symmetric about the qubit step (see also Fig.~\ref{fig:3}A). For amplitudes $A^*/2 \leq A \leq A^*$, the $\Delta_{0,1}$ ($\Delta_{1,0}$) side avoided crossing dominates the dynamics, resulting in a second pair of thresholds $A = A^*-|\delta f_{\rm {dc}}|$, which define the back side of the diamond. In the region outside of the diamond's backside, the qubit is cooled. As the diamond narrows to the point $A = A^*$, the narrowest qubit step is observed. This is the universal cooling condition: only one of the two side avoided crossings ($\Delta_{0,1}$ 
or $\Delta_{1,0}$) is reached and, thereby, strong transitions with relaxation to the ground state result for all $\delta f_{\rm dc}$. In contrast, for $A > A^*$, both side avoided crossings ($\Delta_{0,1}$ 
and $\Delta_{1,0}$) are reached simultaneously for $|\delta f_{\rm{dc}}| < A-A^*$, leading once again to a large population transfer between $|0R\rangle$ and $|0L\rangle$, and opening the second diamond feature (see Fig.~\ref{fig:5}).

The cooling exhibits a rich structure as a function of driving frequency and detuning, resulting from the manner in which state $|1R\rangle$ is accessed (Fig. \ref{fig:4}C). Transitions occur via a (multiphoton) resonant or adiabatic passage process when the driving frequency is high or low enough, respectively~\cite{Berns06a,Valenzuela06a}. At high frequencies (800 and 400 MHz in Fig.~\ref{fig:4}E) well-resolved resonances of n-photon transitions are observed and cooling is thus maximized near resonances. At intermediate frequencies (400 and 200 MHz), Mach-Zehnder interference at the side avoided crossing $\Delta_{01}$ becomes more prominent and modulates the intensity of the n-photon resonances~\cite{Oliver05a,Berns06a}. Below $\nu=200\,{\rm MHz}$, individual resonances are no longer discernible, but as in Fig.~\ref{fig:3}C, the modulation envelope persists~\cite{Berns06a}. At the lowest frequencies ($\nu < 10\,{\rm MHz}$), state $|1R\rangle$ is reached via adiabatic passage through the $\Delta_{01}$ crossing (Fig.~\ref{fig:4}C), and the population transfer and cooling become conveniently independent of detuning (see $\nu=5 ~ {\rm MHz}$ in Fig.~\ref{fig:4}E). As shown in Fig. \ref{fig:4}E, we achieve an effective qubit temperature $T_{\rm eff}$ = 3 mK, even for $T_{\rm bath}$ = 300 mK. In our qubit, our determination of $T_{\rm eff}$ was limited primarily by decoherence (linewidth), which limited the resolution with which we could distinguish the states $|0R\rangle$ and $|0L\rangle$ near degeneracy. Nonetheless, we can estimate the ideally resolvable cooling factor $\alpha_{\rm c}$ for this type of cooling process using Eq.~\ref{Boltzmann},
\begin{equation}
 \alpha_{\rm c} \equiv \frac{T_{\rm bath}}{T_{\rm eff}} =
 \frac{\varepsilon_{1R \to 0R}}{\Delta},
\end{equation}
where $\varepsilon_{1R \to 0R} \approx h \times 25$~GHz is the energy separation where the relaxation $|1R\rangle \to |0R\rangle$ occurs and $\Delta \approx h \times 0.01$~GHz for our qubit, yielding a cooling factor $\alpha_{\rm c} \sim 2500$. For a bath temperature $T_{\rm bath} = 50$~mK, this would correspond to an effective temperature $T_{\rm eff} = 20$~$\mu$K in our qubit. 

Cooling a qubit in equilibrium with the bath requires a characteristic cooling time. In turn, a cooled qubit will thermalize to the environmental bath temperature over a characteristic equilibration time. The relationship between these two times determines if it is possible to drive the qubit while it is still cold. We found in this qubit that equilibration times are at least one order of magnitude larger than cooling times at $T_{\rm bath} < 250$ mK and up to three order of magnitude larger at $T_{\rm bath} < 100$ mK~\cite{Valenzuela06a}. This allowed us ample time to drive the qubit after cooling it.
The implementation of an active cooling pulse prior to a generic driving pulse is highly advantageous.
On the one hand, it sensibly shortens measurement times, enabling us to acquire data at repetition rates that far exceed the intrinsic equilibration rate due to interwell relaxation after each measurement trial. By adopting active cooling, we gained a factor 50 in data acquisition speed, limited by the bandwidth of our readout circuit. On the other hand, by analogy to the cooling of trapped ions and atoms, active cooling greatly reduces
thermal smearing, allowing us to analyze features in the data that would have been hidden otherwise. 
This type of active cooling protocol was required to obtain the detailed, clean data in Fig.~\ref{fig:3}, where we could clearly resolve resonances separated by only 270 MHz (13 mK). In fact, the necessity of active cooling becomes even more evident in the next section where, without the cooling pulse, the observed level of detail could not have been resolved over such a large parameter space in practical acquisition times.

%

%
\section{Amplitude Spectroscopy}
\label{sec:AS}

Frequency-dependent absorption and emission spectroscopy has long played a fundamental role in the characterization of quantum systems. The development of coherent microwave (maser) and optical (laser) sources, high-intensity radiation with tunable, narrow spectral linewidth, has further enabled  targeted absorption spectroscopy of atoms and molecules with high frequency resolution~\cite{Schawlow82a,Thompson85a}. However, the application of broadband frequency spectroscopy is not universally straightforward. This is particularly relevant for artificial atoms engineered from solid-state materials which, when cooled to cryogenic temperatures, assume quantized energy levels that extend into microwave and millimeter wave regimes (10-300 GHz). Although certainly not an impossible task, a broadband frequency-based spectroscopic study of these atoms in excess of around 50 GHz becomes extremely challenging and expensive to implement due to numerous frequency-dependent effects ({\em e.g.}, frequency dispersion and the requisite tolerances to control impedance).

Amplitude spectroscopy is a technique that allows broadband spectroscopic characterization of a quantum system. With amplitude spectroscopy, spectroscopic information is obtained from the system response to driving-field amplitude at a {\em fixed} frequency. The resulting spectroscopic interference patterns, ``spectroscopy diamonds,''  are mediated by multilevel LZS transitions and Mach-Zehnder-type interferometry, and they serve as a fingerprint of the artificial atom's multilevel energy spectrum (Figs.~\ref{fig:5}A and~\ref{fig:5}B). The energy spectrum is then determined by analyzing the atomic fingerprint. In this way, the amplitude spectroscopy technique complements frequency spectroscopy: although a less direct approach, it allows one to probe the energy level structure of a quantum system over extraordinarily large (even practically prohibitive) bandwidths by circumventing many of the challenges associated with a frequency-based approach.

In general, the spectroscopy diamonds arise due to an interplay between the static flux detuning $\delta f_{\rm {dc}}$ and driving amplitude $V$. As described in Sections~\ref{sec:LZS} and~\ref{sec:MZ}, transitions occur when an avoided crossing is reached for a particular set of values ($\delta f_{\rm {dc}}$,$V$). For example, at a flux detuning  $\delta f^*_{\rm {dc}}$, the diamond boundaries occur at $V = V_1, V_2, V_3....$ (Fig.~\ref{fig:5} A and C). The Mach-Zehnder interference due to a phase accumulation $\Delta \theta_{q,q'}$ at a given avoided crossing $\Delta_{q,q'}$ can be modulated by varying both $\delta f_{\rm {dc}}$ and $V$.

There are two important contributions to the diamond spectroscopy patterns: LZS-mediated transitions and intrawell relaxation. In Fig.~\ref{fig:5}D, we show a subset of the diamond interference pattern in Fig.~\ref{fig:5}A. Arrows indicate lines of constant phase accumulation $\Delta \theta_{q,q'}=(2N-1)\pi$ in ($\delta f_{\rm {dc}}$,$V$)-space that leads to LZS transitions at each of the three listed avoided crossings, $\Delta_{0,0}$, $\Delta_{0,1}$, and $\Delta_{1,0}$. Where these lines cross, competition (coordination) between avoided crossings act to suppress (enhance) the net transition rate between pairs of energy levels. The arrangement of these crossing lines leads to the checkerboard patterns observed inside and outside the diamonds.

The second contribution, intrawell relaxation, provides another means to connect energy levels and results in both cooling and population inversion. In Section~\ref{sec:Cooling}, Fig.~\ref{fig:4}C, $\Delta_{0,1}$ mediated the transition $|0L\rangle \to |1R\rangle$, and intrawell relaxation then mediated the transition $|1R\rangle \to |0R\rangle$; the net result was cooling, since the flux $\delta f_{\rm {dc}}$ was positive, making $|0R\rangle$ the ground state. However, in Fig.~\ref{fig:5}D $\delta f_{\rm {dc}}$ is negative. Furthermore, in the upper-right corner of Fig.~\ref{fig:5}D, 
both crossings $\Delta_{0,1}$ and $\Delta_{1,0}$ are accessed. In the bright red regions, $\Delta_{0,1}$ still tends to cause transitions $|0L\rangle \to |1R\rangle$, and relaxation puts that population in state $|0R\rangle$. However, the interference condition at $\Delta_{1,0}$ on the other side of the energy level diagram (Fig.~\ref{fig:5}C) tends to keep the population in $|0R\rangle$. Therefore, population builds in state $|0R\rangle$, the first excited state for negative flux detuning, resulting in strong population inversion. Varying the interference conditions at $\Delta_{0,1}$ and $\Delta_{1,0}$ by changing ($\delta f_{\rm {dc}}$,$V$) causes the observed modulation between population inversion and cooling.

We have developed several techniques for extracting information about the energy levels from the spectroscopy interference patterns~\cite{Berns08a,Rudner08a}. The key metrics are the positions of the avoided crossings in flux, the values of the splittings $\Delta_{q,q'}$, and the slopes of the energy levels. With this information, one can reconstruct a large portion of the energy level diagram.

The positions of the avoided crossings can be determined precisely from the diamond boundaries, because the onset of each diamond is associated with LZS transitions at a particular level crossing. The splitting of each avoided crossing can be obtained essentially by fitting the LZS formula in Eq.~\ref{Eq:P_LZS} to the Mach-Zehnder interference patterns. Alternatively, one can study the dynamical population transfer between states using the pulsed, short-time implementation of amplitude spectroscopy~\cite{Berns08a}.

The energy-level slopes can be determined by two means. The first is by relating the separation between Mach-Zehnder interference nodes to the expected phase accumulation $\Delta \theta_{12}$, which depends sensitively on the energy-level slope. Alternatively, we show with Rudner {\em et al.} in Ref.~\onlinecite{Rudner08a} that the two-dimensional Fourier transform of the diamond patterns yields a family of {\em one-dimensional} sinusoids in Fourier space; the periods of these sinusoids are related to the energy-level slopes. The intuition is that the Fourier transform inverts the energy domain of the spectroscopy to the time domain (scaled by $\hbar$). This means that the sinusoids in Fourier space image the time-dependence of the quantum phase of the qubit, which, in turn, depends sensitively on the energy-level slopes.

Using amplitude spectroscopy, we were able to scan the energy level diagram continuously beyond the fourth energy-level avoided crossing ($\Delta_{0,0}\ldots\Delta_{0,4}$, $\Delta_{4,0}$) with splitting values ranging from $\Delta_{0,0}\approx 0.01$~GHz to $\Delta_{0,4}, \Delta_{4,0}\approx 2.2$~GHz~\cite{Berns08a}. The equivalent information obtained using frequency spectroscopy would have required scanning frequencies from 0.01 GHz out to beyond 100 GHz (in this device, each avoided crossing is separated by approximately 25 GHz, and so each subsequent crossing in Fig.~\ref{fig:5} is raised an additional 25 GHz above the ground state). Remarkably, with amplitude spectroscopy, the entire scan performed in Fig.~\ref{fig:5}A was performed at a fixed frequency $\nu=185$~MHz. The scan in Fig.~\ref{fig:5}B shows the amplitude spectroscopy of the same system for a fixed frequency $\nu=700$~MHz, clearly in the discrete resonance limit. The resonance condition adds another constraint, making a more complex and interesting checkerboard pattern.

\section{Summary and Conclusions}
\label{sec:Summary}
Strongly driving a superconducting artificial atom through an avoided level crossing results in a Landau-Zener-St\"{u}ckelberg transition, which, in general, creates a superposition of atomic states whose weighting depends on the size of the avoided crossing and the velocity with which it is traversed. In this sense, as we discussed in Section~\ref{sec:LZS}, the LZS mechanism acts as a beamsplitter for artificial atoms.

In Section~\ref{sec:MZ}, we described how harmonically driving the system cascades two LZS transitions per driving period, resulting in an atomic analog to a Mach-Zehnder interferometer. The relative phase acquired between LZS transitions is the interference phase of the interferometer, and it is tunable by the driving amplitude. The buildup of population over many driving periods exhibits St\"{u}ckelberg oscillations as a function of the driving amplitude (interference phase) due to the cascaded Mach-Zehnder-type interference effect. To observe these oscillations, the coherence time must only be longer than the small portion of the drive period during which the interference phase accrues, and the relaxation time must be long enough to maintain the population until readout.

The total phase accumulated over one period, in contrast, is amplitude independent. For coherence times longer than the drive period, cascaded interferometers yield a net population change when this round-trip phase accumulation is $2 \pi n$ per driving period, a condition which can be viewed as the ``time-domain'' counterpart to the $n$-photon resonance condition $n h \nu = \varepsilon_{0,n}$. By making the drive period commensurate with the coherence time, we showed that we could still observe the St\"{u}ckelberg oscillations, even though the discrete resonances were no longer discernable.

We utilized strong driving and the LZS mechanism with higher-energy levels to achieve both cooling and population inversion in our artificial atom. In Section~\ref{sec:Cooling}, we described using a microwave pumping scheme to cool the atomic degrees of freedom a factor 10-100 times colder than the ambient dilution refrigerator temperature.  The scheme involved pumping unwanted thermal population to an ancillary excited state, from which it relaxed to the ground state. In Section ~\ref{sec:AS}, we showed that by reversing the order, we could pump population through an ancillary state to achieve inversion.

The energy level structure can be probed over extraordinarily large bandwidth using the amplitude spectroscopy approach presented in Section~\ref{sec:AS}. Since the LZS mechanism and Mach-Zehnder interference are sensitive to the defining features in the energy level diagram (energy band slopes, level splittings, and their locations), the interference patterns that result from sweeping the amplitude are a ``fingerprint'' of the artificial atom's energy spectrum. Using amplitude spectroscopy at a fixed driving frequency of only 185 MHz, we could access continuously multiple energy levels from about 10 MHz out to beyond 120 GHz.

Large-amplitude driving and the LZS mechanism have application to quantum information science and technology. Active cooling has utility in state initialization and refreshing ancilla qubits in quantum error correction protocols. Amplitude spectroscopy provides a means to ascertain over large bandwidth the energy level structure of a qubit system beyond the lowest two levels. The Mach-Zehnder-type interference can facilitate nonadiabatic control schemes, in which the quantum interference at an avoided crossing is used to achieve state transitions that approach the intrinsic coupling rate $\Delta$. In cold atoms and molecules, this kind of non-adiabatic control has been used to drive transitions that would otherwise be challenging to realize in a weak-driving limit.

We thank T.P. Orlando, D.M. Berns, L.S. Levitov, M.S. Rudner, A.V. Shytov, and K.K. Berggren with whom we collaborated closely on this work; J. Bylander, S. Gustavsson, B. Turek, and J. Sage for carefully reading the manuscript; V. Bolkhovsky, G. Fitch, T.R Jordan, E. Macedo, P. Murphy, K. Parrillo, R. Slattery, and T. Weir at MIT Lincoln Laboratory for technical assistance.
This work was supported by AFOSR and LPS (F49620-01-1-0457) under the DURINT program, and by the U.S. Government. The work at Lincoln Laboratory was sponsored by the US DoD under Air Force Contract No. FA8721-05-C-0002.


\begin{thebibliography}{}
%
%



\bibitem{Clarke88}
J. Clarke, A. N. Cleland, M. H. Devoret, D. Esteve, J. H. Martinis,
Quantum mechanics of a macroscopic variable: the phase difference of a Josephson junction,
\emph{Science} \textbf{239}, 992-997 (1988).

\bibitem{Clarke08a}
J. Clarke and F.K. Wilhelm,
Superconducting quantum bits,
\emph{Nature} \textbf{453}, 1031-1042 (2008).


\bibitem{Mooij99a}
J.~E.~Mooij, T.~P.~Orlando, L.~S.~Levitov, L.~Tian, C.~H.~{van~der~Wal},
 S.~Lloyd,
Josephson persistent-current qubit,
\emph{Science} \textbf{285},
1036-1039 (1999).

\bibitem{Orlando99a}
T.~P.~Orlando, J.~E.~Mooij, L.~Tian, C.~H.~{van~der~Wal},
L.~S.~Levitov, S.~Lloyd, and J.~J.~Mazo,
Superconducting persistent-current qubit,
\emph{Phys. Rev.~B} \textbf{60},
15398-15413 (1999).


\bibitem{Oliver05a}
W. D. Oliver, Y. Yu, J. C. Lee, K. K. Berggren, L. S. Levitov, T. P. Orlando,
Mach-Zehnder interferometry in a strongly driven superconducting qubit,
\emph{Science} 310, 1653-1657 (2005).

\bibitem{Berns06a}
D.M. Berns, W.D. Oliver, S.O. Valenzuela, A.V. Shytov, K.K. Berggren, L.S. Levitov, T.P. Orlando,
Coherent Quasiclassical Dynamics of a Persistent Current Qubit,
\emph{Phys. Rev. Lett.} 97, 150502 (2006).

\bibitem{Valenzuela06a}
S. O. Valenzuela, W. D. Oliver, D. M. Berns, K. K. Berggren, L. S. Levitov, T. P. Orlando,
Microwave-induced cooling of a superconducting qubit,
\emph{Science} 314, 1589-1592 (2006).

\bibitem{Berns08a}
D.M. Berns, M.S. Rudner, S.O. Valenzuela, K. K. Berggren, W. D. Oliver, L. S. Levitov, T. P. Orlando,
Amplitude spectroscopy of a solid-state artificial atom,
\emph{Nature} 455, 51-57 (2008).

\bibitem{Rudner08a}
M.S. Rudner, A.V. Shytov, L. S. Levitov, D. M. Berns, W. D. Oliver, S. O. Valenzuela, T. P. Orlando,
Quantum phase tomography of a strongly driven qubit,
\emph{Phys. Rev. Lett.} 101, 190502 (2008).




\bibitem{Friedman00a}
J.~R. Friedman,
V. Patel, W. Chen, S. K. Tolpygo, J. E. Lukens,
Quantum superposition of distinct macroscopic states,
\emph{Nature} \textbf{406}, 43-46 (2000).

\bibitem{Wal00a}
C.~H. van der Wal, A. C. J. ter Haar, F. K. Wilhelm, R. N. Schouten,
C. J. P. M. Harmans, T. P. Orlando, S. Lloyd, J. E. Mooij,
Quantum superposition of macroscopic persistent-current states,
\emph{Science} \textbf{290}, 773-777 (2000).

\bibitem{Berkley03a}
A.~J. Berkley, H. Xu, R. C. Ramos, M. A. Gubrud, F. W. Strauch,
P. R. Johnson, J. R. Anderson, A. J. Dragt, C. J. Lobb, F. C. Wellstood,
Entangled macroscopic quantum states in two superconducting qubits,
\emph{Science} \textbf{300}, 1548-1550 (2003).


\bibitem{Nakamura99a}
Y.~Nakamura, Y.~A. Pashkin, J.~S. Tsai,
Coherent control of macroscopic quantum states in a single-Cooper-pair box,
\emph{Nature} \textbf{398}, 786-788 (1999).

\bibitem{Nakamura01}
Y.~Nakamura, Y.~A. Pashkin, J.~S. Tsai,
Rabi oscillations in a large Josephson-junction charge two-level system,
\emph{Phys. Rev. Lett.} \textbf{87}, 246601 (2001).

\bibitem{Vion02a}
D.~Vion, A. Aassime, A. Cottet, P. Joyez, H. Pothier, C. Urbina, D. Esteve, and M. H. Devoret,
Manipulating the quantum state of an electrical circuit,
\emph{Science} \textbf{296}, 886-889 (2002).

\bibitem{Yu02a}
Y.~Yu, S. Han, X. Chu, S.-I. Chu, Z. Wang,
Coherent temporal oscillations of macroscopic quantum states in a Josephson junction,
\emph{Science} \textbf{296}, 889-892 (2002).

\bibitem{Martinis02a}
J.~M. Martinis, S.~Nam, J.~Aumentado, C.~Urbina,
Rabi oscillations in a large Josephson-junction qubit,
\emph{Phys. Rev. Lett.} \textbf{89}, 117901 (2002).

\bibitem{Chiorescu03a}
I.~Chiorescu, Y.~Nakamura, C.~J. P.~M. Harmans, J.~E. Mooij,
Coherent quantum dynamics of a superconducting flux qubit,
\emph{Science} \textbf{299}, 1869-1871 (2003).



\bibitem{Claudon04a}
J.~Claudon, F.~Balestro, F.~W.~J. Hekking, O.~Buisson,
Coherent oscillations in a superconducting multilevel quantum system,
\emph{Phys. Rev. Lett.}
\textbf{93}, 187003 (2004).

\bibitem{Plourde05a}
B. L. T. Plourde, T.L. Robertson, P.A. Reichardt, T. Hime, S. Linzen, C.-E. Wu, and John Clarke,
Flux qubits and readout device with two independent flux lines,
\emph{Phys. Rev. B}
\textbf{72}, 060506(R) (2005).

\bibitem{Saito06a}
S.~Saito,
T. Meno, M. Ueda, H. Tanaka, K. Semba, and H. Takayanagi,
Parametric control of a superconducting flux qubit,
\emph{Phys. Rev. Lett.} \textbf{96}, 107001 (2006).


\bibitem{Lisenfeld07a}
J. Lisenfeld, A. Lukashenko, M. Ansmann, J.M. Martinis, and A.V. Ustinov,
Temperature dependence of coherent oscillations in josephson phase qubits,
\emph{Phys. Rev. Lett.} \textbf{99}, 170504 (2007).


\bibitem{Izmalkov04a}
A. Izmalkov, M. Grajcar, E. Il'ichev, N. Oukhanski, Th. Wagner, H.-G. Meyer, W. Krech, M. H. S. Amin, A. Maassen van den Brink and A. M. Zagoskin,
Observation of macroscopic Landau-Zener transitions in a superconducting device,
\emph{Europhys. Lett.} 65, 844-849 (2004).


\bibitem{Sillanpaa06a}
M. Sillanpaa, T. Lehtinen, A. Paila, Yu. Makhlin, P. Hakonen,
Continuous-time monitoring of Landau-Zener interference in a Cooper-pair box,
\emph{Phys. Rev. Lett.} 96, 187002 (2006).


\bibitem{Wilson07a}
C.M. Wilson, T. Duty, F. Persson, M. Sandberg, G. Johansson, and P. Delsing,
Coherent times of dressed states of a superconducting qubit under extreme driving,
\emph{Phys. Rev. Lett.} 98, 257003 (2007).

\bibitem{Izmalkov08a}
A. Izmalkov, S.J.W. van der Ploeg, S.N. Shevchenko, M. Grajcar, E. Il'ichev, U. H\"{u}bner, A.N. Omelyanchouk, and H.-G.Meyer,
Consistency of ground state and spectroscopic measurements on flux qubits,
\emph{Phys. Rev. Lett.} 101, 017003 (2008).



\bibitem{Niskanen07b}
A.O. Niskanen, Y. Nakamura, and J.P. Pekola,
Information entropic superconducting microcooler,
\emph{Phys. Rev. B} {\bf 76}, 174523 (2007).

\bibitem{You08a}
J.Q. You, Yu-xi Liu, and Franco Nori,
Simultaneous cooling of an artificial atom and its neighboring quantum system,
\emph{Phys. Rev. Lett.} 100, 047001 (2008).

\bibitem{Grajcar08a}
M. Grajcar, S.H.W. van der Ploeg, A. Izmalkov, E. Il'ichev, H.-G. Meyer, A. Fedorov, A. Shnirman, and G. Sch\"{o}n,
Sisyphus cooling and amplification by a superconducting qubit,
\emph{Nature Phys.} 4, 612-616 (2008).

\bibitem{Murali04a}
K.V.R.M. Murali, Z. Dutton, W.D. Oliver, D.S. Crankshaw, and T.P. Orlando,
Probing decoherence with electromagnetically induced transparency in superconductive quantum circuits,
\emph{Phys. Rev. Lett.} \textbf{93}, 087003 (2004).

\bibitem{Dutton06a}
Z. Dutton, K.V.R.M. Murali, W.D. Oliver, and T.P. Orlando,
Electromagnetically induced transparency in superconducting quantum circuits: effects of decoherence, tunneling, and multilevel crosstalk,
\emph{Phys. Rev. B} \textbf{73}, 104516 (2006).


\bibitem{Leek07a}
P.J. Leek, J.M. Fink, A. Blais, R. Bianchetti, M. Goppl, J.M. Gambetta, D.I. Schuster, L. Frunzio, R.J. Schoelkopf, and A. Wallraff,
Observation of Berry's phase in a solid-state qubit,
\emph{Science} {\bf 318}, 1889-1892 (2007).



\bibitem{Chiorescu04a}
I. Chiorescu,
P. Bertet, K. Semba, Y. Nakamura, C.J.P.M. Harmans, and J.E. Mooij,
Coherent dynamics of a flux qubit coupled to a harmonic oscillator,
\emph{Nature} {\bf 431}, 159-162 (2004).

\bibitem{Wallraff04a}
A. Wallraff,
D. I. Schuster, A. Blais, L. Frunzio, R.-S. Huang,
J. Majer, S. Kumar, S. M. Girvin and R. J. Schoelkopf,
Strong coupling of a single photon to a superconducting qubit using circuit quantum electrodynamics,
\emph{Nature} {\bf 431}, 162-167 (2004).

\bibitem{Johansson06a}
J.~Johansson,
S. Saito, T. Meno, H. Nakano, M. Ueda, H. Tanaka, K. Semba, and H. Takayanagi,
Vacuum rabi oscillations in a macroscopic superconducting qubit LC oscillator system,
\emph{Phys. Rev. Lett.} \textbf{96}, 127006 (2006).

\bibitem{Deppe08a}
F. Deppe, M. Mariantoni, E.P. Menzel, A. Marx, S. Saito, K. Kakuyanagi, H. Tanaka, T. Menon, K. Semba, H. Takayanagi, E. Solano, and R. Gross
Two-photon probe of the Jaynes-Cummings model and controlled symmetry breaking in circuit QED,
\emph{Nature Phys.} {\bf 4}, 686-691 (2008).

\bibitem{Fink08a}
J.M. Fink, M. Goppl, M. Baur, R. Bianchetti, P.J. Leek, A. Blais, and A. Wallraff,
Climbing the Jaynes-Cummings ladder and observing its root n nonlinearity in a cavity QED system,
\emph{Nature} {\bf 454}, 315-318 (2008).

\bibitem{Fragner08a}
A. Fragner, M. Goppl, J.M. Fink, M. Baur, R. Bianchetti, P.J. Leek, A. Blais, and A. Wallraff,
Resolving vacuum fluctuations in an electrical circuit by measuring the Lamb shift,
\emph{Science} {\bf 322}, 1357-1360 (2008).

\bibitem{Hofheinz08a}
M. Hofheinz, E.M. Weig, M. Ansmann, R.C. Bialczak, E. Lucero, M. Neeley, A.D. O'Connell, H. Wang, J.M. Martinis, and A.N. Cleland,
Generation of Fock states in a superconducting quantum circuit,
\emph{Nature} {\bf 454}, 310-314 (2008).


\bibitem{Makhlin01a}
Y.~Makhlin, G.~Sch\"on, A.~Shnirman,
Quantum-state engineering with Josephson-junction devices,
\emph{Rev. Mod. Phys.} \textbf{73}, 357-400 (2001).

\bibitem{Mooij05}
J. E. Mooij, The road to quantum computing,
 \emph{Science} {\bf 307}, 1210-1211 (2005)


\bibitem{Hime06a} T. Hime, 
    P.A. Reichart, B.L.T. Plourde, T.L. Robertson, C.-E. Wu, A.V. Ustinov, and J. Clarke,
    Solid-state qubits with current-controlled coupling,
    \emph{Science} {\bf 314} 1427-1429 (2006)

\bibitem{Ploeg07a}
    S.H.W. van der Ploeg, 
    A. Izmalkov, A. Maassen van den Brink, U. H\"{u}bner, M.
    Grajcar, E. Il'ichev, H.-G. Meyer, and A.M. Zagoskin,
    Controllable coupling of superconducting flux qubits,
    \emph{Phys. Rev. Lett.} {\bf 98}, 057004 (2007).

\bibitem{Niskanen07a}
    A.O. Niskanen, K. Harrabi, F. Yoshihara, Y. Nakamura, S. Lloyd, and J.S. Tsai,
    Quantum coherent tunable coupling of superconducting qubits,
    \emph{Science} {\bf 316}, 723-726 (2007).

\bibitem{Kerman08a}
    A.J. Kerman and W.D. Oliver,
    High-fidelity quantum operations on superconducting qubits in the presence of noise,
    \emph{Phys. Rev. Lett.} {\bf 101}, 070501 (2008).


\bibitem{Sillanpaa07a}
M.A. Sillanpaa, J.I. Park, and R.W. Simmonds,
Coherent quantum state storage and transfer between two phase qubits via a resonant cavity,
\emph{Nature} {\bf 449}, 438-442 (2007).

\bibitem{Majer07a}
J. Majer, J.M. Chow, J.M. Gambetta, J. Koch, B.R. Johnson, J.A. Schreier, L. Frunzio, D.I. Schuster,
A.A. Houck, A. Wallraff, A. Blais, M.H. Devoret, S.M. Girvin, and R.J. Schoelkopf
Coupling superconducting qubits via a cavity bus,
\emph{Nature} {\bf 449}, 443-447 (2007).


\bibitem{Pashkin03a}
Y.~A. Pashkin, T. Yamamoto, O. Astafiev, Y. Nakamura, D. V. Averin, and J. S. Tsai,
Quantum oscillations in two coupled charge qubits,
\emph{Nature} \textbf{421}, 823-826 (2003).

\bibitem{Yamamoto03a}
T.~Yamamoto, Yu. A. Pashkin, O. Astafiev, Y. Nakamura, J. S. Tsai,
Demonstration of conditional gate operation using superconducting charge qubits,
\emph{Nature} \textbf{425}, 941-944 (2003).

\bibitem{McDermott05a}
R.~McDermott, R. W. Simmonds, M. Steffen, K. B. Cooper, K. Cicak,
K. D. Osborn, S. Oh, D. P. Pappas, J. M. Martinis,
Simultaneous state measurement of coupled Josephson phase qubits,
\emph{Science} \textbf{307}, 1299-1302 (2005).

\bibitem{Plantenberg07a}
J.H. Plantenberg, P.C. de Groot, C.J.P.M. Harmans, and J.E. Mooij,
Demonstration of controlled-NOT quantum gates on a pair of superconducting quantum bits,
\emph{Nature} \textbf{447}, 836-839 (2007).


\bibitem{Steffen06a}
M. Steffen, A. Ansmann, R.C. Bialczak, N. Katz, E. Lucero, R.~McDermott, M. Neeley, E.M. Weig, A.N. Cleland, and J. M. Martinis,
Measurement of the entanglement of two superconducting qubits via state tomography,
\emph{Science} \textbf{313}, 1423-1425 (2006).

\bibitem{Steffen06b}
M. Steffen, A. Ansmann, R.~McDermott, N. Katz, R.C. Bialczak, E. Lucero, M. Neeley, E.M. Weig, A.N. Cleland, and J. M. Martinis,
State tomography of capacitively shunted phase qubits with high fidelity,
\emph{Phys. Rev. Lett.} \textbf{97}, 050502 (2006).


\bibitem{Neeley08a}
M. Neeley, M. Ansmann, R.C. Bialczak, M. Hofheinz, N. Katz, E. Lucero, A. O'Connell, H. Wang, A.N. Cleland, and J. M. Martinis,
Process tomography of quantum memory in a Josephson-phase qubit coupled to a two-level state,
\emph{Nature Phys.} \textbf{4}, 523-526 (2008).


\bibitem{Siddiqi04a}
I. Siddiqi, 
\emph{et~al.},
RF-driven Josephson bifurcation amplifier for quantum measurement,
\emph{Phys. Rev. Lett.} {\bf 93}, 207002 (2004)

\bibitem{Katz06a}
N. Katz, M. Ansmann, R.C. Bialczak, E. Lucero, R. McDermott, M. Neeley, M. Steffen,
E.M. Weig, A.N. Cleland, J.M. Martinis, A.N. Korotkov
Coherent state evolution in a superconducting qubit from partial-collapse measurement,
\emph{Science} {\bf 312}, 1498-1500 (2006)

\bibitem{Lupascu07a}
A. Lupascu, S. Saito, T. Picot, P.C. de Groot, C.J.P.M. Harmans, J.E. Mooij,
Quantum non-demolition measurement of a superconducting two-level system,
\emph{Nature Physics} {\bf 3}, 119 (2007)

\bibitem{Yamamoto08a}
T. Yamamoto, K. Inomata, M. Watanabe, K. Matsuba, T. Miyazaki, W.D. Oliver, Y. Nakamura, and J.S. Tsai,
Flux-driven Josephson parametric amplifier,
\emph{Appl. Phys. Lett.} {\bf 93}, 042510 (2008)

\bibitem{Beltran08a}
M.A. Castellanos-Beltran, K.D. Irwin, G.C. Hilton, L.R. Vale, and K.W. Lehnert,
Amplification and squeezing of quantum noise with a tunable Josephson metamaterial,
\emph{Nature Physics} {\bf 4}, 929-931 (2008)

\bibitem{Naaman08a}
O. Naaman, J. Aumentado, L. Friedland, J.S. Wurtele, I. Siddiqi,
Phase-locking transition in a chirped superconducting Josephson resonator,
\emph{Phys. Rev. Lett.} {\bf 101}, 117005 (2008)


\bibitem{Lee05a}
J.C. Lee, W.D. Oliver, T.P. Orlando, and K.K. Berggren,
Resonant readout of a persistent current qubit,
\emph{IEEE Trans. Appl. Supercond.} 15, 841-844 (2005).

\bibitem{Lee07a}
J.C. Lee, W.D. Oliver, K.K. Berggren, and T.P. Orlando,
Nonlinear resonant behavior of a dispersive readout circuit for a superconducting flux qubit,
\emph{Phys. Rev. B} 75, 144505 (2007).




\bibitem{Landau32a}
L.D. Landau, \emph{Phys. Z. USSR} {\bf 2}, 46-51
On the theory of transfer of energy at collisions II,
(1932).

\bibitem{Zener32a}
C. Zener,
Non-adiabatic crossing of energy levels,
\emph{Proc. R. Soc. Lond. A} {\bf 137}, 696-702
(1932).

\bibitem{Stueckelberg32a}
E. C. G. Stueckelberg,
Theorie der un elastischen St\"{o}sse zwischen Atomen,
\emph{Helv. Phys. Acta} {\bf 5}, 369-422
(1932).



\bibitem{H_Nakamura01a}
H. Nakamura, \emph{Nonadiabatic Transition} (London, England:
World Scientific, 2001).



\bibitem{Grifoni98a}
M. Grifoni and P. H\"{a}nggi,
Driven quantum tunneling,
\emph{Phys. Rep.} {\bf 304}, 229-354 (1998).


\bibitem{Kayanuma93a}
Y. Kayanuma,
Phase coherence and nonadiabatic transition at a level crossing in a periodically driven two-level system,
\emph{Phys. Rev. B} {\bf 47}, 9940-9943 (1993).

\bibitem{Kayanuma94a}
Y. Kayanuma,
Role of phase coherence in the transition dynamics of a periodically driven two-level system,
\emph{Phys. Rev. B} {\bf 50}, 843-845 (1994).



\bibitem{Kayanuma97a}
Y. Kayanuma,
Stokes phase and geometrical phase in a driven two-level system,
\emph{Phys. Rev. A} {\bf 55}, R2495-R2498 (1997).

\bibitem{Kayanuma00a}
Y. Kayanuma and Y. Mizumoto,
Landau-Zener transitions in a level-crossing system with periodic modulation of the diagonal energy,
\emph{Phys. Rev. A} {\bf 62}, 061401(R) (2000).


\bibitem{Shytov03}
A. V. Shytov, D. A. Ivanov, M. V. Feigel'man,
Landau-Zener interferometry for qubits,
\emph{Eur. Phys. J. B} {\bf 36}, 263-269 (2003)



\bibitem{Ashab07a}
S. Ashhab, J.R. Johansson, A.M. Zagoskin, and Franco Nori,
Two-level systems driven by large-amplitude fields,
\emph{Phys. Rev. A} {\bf 75}, 063414 (2007).

\bibitem{Wubs06a}
M. Wubs, K. Saito, S. Kohler, P. H\"{a}nggi, and Y. Kayanuma,
Gauging a quantum heat bath with dissipative Landau-Zener Transitions,
\emph{Phys. Rev. Lett.} {\bf 97}, 200404 (2006).

\bibitem{KSaito06a}
K. Saito, M. Wubs, S. Kohler, P. H\"{a}nggi, and Y. Kayanuma,
Quantum state preparation in circuit QED via Landau-Zener tunneling,
\emph{Europhys. Lett.} {\bf 76}, 22-28 (2006).

\bibitem{KSaito07a}
K. Saito, M. Wubs, S. Kohler, Y. Kayanuma, and P. H\"{a}nggi,
Dissipative Landau-Zener transitions of a qubit: bath-specific and universal behavior,
\emph{Phys. Rev. B} {\bf 75}, 214308 (2007).


\bibitem{Grossmann91a}
F. Grossmann, T. Dittrich, P. Jung, and P. H\"{a}nggi
Coherent destruction of tunneling,
\emph{Phys. Rev. Lett.} {\bf 67}, 516-519 (1991).

\bibitem{Llorente92a}
J.M. Gomez Llorente and J. Plata,
Tunneling control in a two-level system,
\emph{Phys. Rev. A} {\bf 45}, R6958-R6961 (1992).

\bibitem{Kayanuma08a}
Y. Kayanuma and K. Saito,
Coherent destruction of tunneling, dynamic localization, and the Landau-Zener formula,
\emph{Phys. Rev. A} {\bf 77}, 010101(R) (2008).

\bibitem{vanderWiel03a}
W. G. van der Wiel,
\emph{et~al.},
Electron transport through double quantum dots,
\emph{Rev. Mod. Phys.} \textbf{75}, 1-22 (2003).

\bibitem{Hanson07a}
R. Hanson, L.P. Kouwenhoven, J.R. Petta, S. Tarucha, and L.M.K. Vandersypen
Spins in few-electron quantum dots,
\emph{Rev. Mod. Phys.} \textbf{79}, 1217-1265 (2007).



\bibitem{Tannoudji92a}
C. Cohen-Tannoudji, J. Dupont-Roc, and G. Grynberg, \emph{Atom-Photon Interactions: Basic Processes and Applications} Ch. 6 (Wiley, 1992).


\bibitem{Friedman96a}
F.R. Friedman, \emph{et~al.},
Macroscopic measurement of resonant magnetization tunnelling in high-spin molecules,
\emph{Phys. Rev. Lett.} {\bf 76}, 3830-3833 (1996).

\bibitem{Thomas96a}
L. Thomas, \emph{et~al.},
Macroscopic quantum tunnelling of magnetization in a single crystal of nanomagnets,
\emph{Nature} {\bf 383}, 145-147 (1996).

\bibitem{Wernsdorfer99a}
W. Wernsdorfer and R. Sessoli,
Quantum phase interference and parity effects in magnetic molecular clusters,
\emph{Science} {\bf 284}, 133-135 (1999).



\bibitem{Baruch92a}
M. C. Baruch, T. F. Gallagher,
Ramsey interference fringes in single pulse microwave multiphoton transitions,
\emph{Phys. Rev. Lett.} {\bf 68}, 3515-3518 (1992).

\bibitem{Yoakum92a}
S. Yoakum, L. Sirko, P. M. Koch,
St\"{u}ckelberg oscillations in the multiphoton excitation of helium Rydberg atoms: Observation with a pulse of coherent, field and suppression by additive noise
\emph{Phys. Rev. Lett.} {\bf 69}, 1919-1922 (1992).


\bibitem{Mark07a}
M. Mark, T. Kraemer, P. Waldburger, J. Herbig, C.Chin, H.-C. N\"{a}gerl, and R. Grimm
``St\"{u}ckelberg interferometry'' with ultracold molecules,
\emph{Phys. Rev. Lett.} {\bf 99}, 113201 (2007).


\bibitem{Mark07b}
M. Mark, F. Ferlaino, S. Knoop, J.G. Danzl, T. Kraemer, C.Chin, H.-C. N\"{a}gerl, and R. Grimm,
Spectroscopy of ultracold trapped cesium Feshbach molecules,
\emph{Phys. Rev. A} {\bf 76}, 042514 (2007).

\bibitem{Lang08a}
F. Lang, P.V.D. Straten, B. Brandst\"{a}tter, G. Thalhammer, K. Winkler, P.S. Julienne, R. Grimm, and J. Hecker Denschlag,
Cruising through molecular bound-state manifolds with radiofrequency,
\emph{Nature Physics} {\bf 4}, 223-226 (2008).

\bibitem{Tien63a}
P.~K. Tien and J. P. Gordon,
Multiphoton process observed in the interaction of microwave fields with the tunneling between superconductor films
\emph{Phys. Rev.} \textbf{129}, 647 (1963).

\bibitem{Kouwenhoven94}
L. P. Kouwenhoven, S. Jauhar, J. Orenstein, P. L. McEuen, Y.
Nagamune, J. Motohisa, and H. Sakaki,
Observation of Photon-Assisted Tunneling through a Quantum Dot,
\emph{Phys. Rev. Lett.} {\bf 73}, 3443 (1994).

\bibitem{NakamuraTsai99}
Y. Nakamura, J. S. Tsai,
A coherent two-level system in a superconducting single-electron transistor observed through photon-assisted cooper-pair tunneling,
\emph{J. Supercond.} {\bf 12}, 799
(1999).



\bibitem{Astafiev07a}
O. Astafiev, K. Inomata, A.O. Niskanen, T. Yamamoto, Yu. A. Pashkin, Y. Nakamura, and J.S. Tsai,
Single artificial-atom lasing,
\emph{Nature} {\bf 449}, 588-590 (2007).


\bibitem{Schawlow82a}
A.L. Schawlow,
Spectroscopy in a new light,
\emph{Rev. Mod. Phys.} \textbf{54}, 697-707 (1982).

\bibitem{Thompson85a}
R.C. Thompson,
High resolution laser spectroscopy of atomic systems,
\emph{Rep. Prog. Phys.} \textbf{48}, 531-578 (1985).







\end{thebibliography}
\end{document}